\documentclass[traditabstract]{aa}
\usepackage{graphicx}
\usepackage{txfonts}
\usepackage{natbib}
\defcitealias{A2005013}{W\&D}
\defcitealias{P2005002}{Paper~I}
\defcitealias{P2006002}{Paper~II}
\bibpunct{(}{)}{;}{a}{}{,} 

\begin{document}

 \title{The distance to the Andromeda Galaxy from eclipsing
binaries\thanks{Based on observations made with the Isaac Newton Telescope
operated on the island of La Palma by the Isaac Newton Group in the Spanish
Observatorio del Roque de los Muchachos of the Instituto de Astrof\'{\i}sica de
Canarias.}\fnmsep\thanks{Based on observations obtained at the Gemini
Observatory, which is operated by the Association of Universities for Research
in As\-tron\-o\-my, Inc., under a cooperative agreement with the NSF on behalf
of the Gemini partnership: the National Science Foundation (U\-nit\-ed States),
the Science and Technology Facilities Council (United Kingdom), the National
Research Council (Canada), CONICYT (Chil\-e), the Australian Research Council
(Australia), Minist\'erio da Ci\^encia e Tecnologia (Brazil) and Ministerio de
Ciencia, Tecnolog\'{\i}a e Innovaci\'on Productiva  (Argentina)}}
 \author{F. Vilardell \inst{1,2} \and
         I. Ribas \inst{3} \and
         C. Jordi \inst{2} \and
         E.L. Fitzpatrick \inst{4} \and 
         E.F. Guinan \inst{4}}

 \institute{Departament de F\'{\i}sica, Enginyeria de Sistemes i Teoria del Senyal, Universitat d'Alacant, Apartat 99, 03080 Alacant, Spain \\
  \email{francesc.vilardell@ua.es}
  \and
  Departament d'Astronomia i Meteorologia (ICC-IEEC), Universitat de Barcelona, c/ Mart\'{\i} i Franqu\`es, 1, 08028 Barcelona, Spain\\
 \email{carme.jordi@am.ub.es}
  \and
  Institut de Ci\`encies de l'Espai (CSIC-IEEC), Campus UAB, Facultat de
Ci\`encies, Torre C5, parell, 2a pl., E-08193 Bellaterra, Spain\\
 \email{iribas@ice.csic.es}
  \and
   Department of Astronomy and Astrophysics, Villanova University, 800 Lancaster Avenue, Villanova, PA 19085, USA\\
 \email{[edward.guinan;edward.fitzpatrick]@villanova.edu}
 }
 
 \date{Received / Accepted}

 \abstract{The cosmic distance scale largely depends on distance determinations
to Local Group galaxies. In this sense, the Andromeda Galaxy (\object{M\,31})
is a key rung to better constrain the cosmic distance ladder. A project was
started in 1999 to firmly establish a direct and accurate distance to M\,31
using eclipsing binaries (EBs). After the determination of the first direct
distance to M\,31 from EBs, the second direct distance to an EB system is
presented: \object{M31V\,J00443610+4129194}. Light and radial velocity curves
were obtained and fitted to derive the masses and radii of the components. The
acquired spectra were combined and disentangled to determine the temperature of
the components. The analysis of the studied EB resulted in a distance
determination to M\,31 of $(m-M)_0=24.30\pm0.11$~mag. This result, when
combined with the previous distance determination to M\,31, results in a
distance modulus of $(m-M)_0=24.36\pm0.08$~mag ($744\pm33$ kpc), fully
compatible with other distance determinations to M\,31. With an error of only
4\%, the obtained value firmly establishes the distance to this important
galaxy and represents the fulfillment of the main goal of our project.}


 \maketitle

\section{Introduction}
\label{introdistm31}

Eclipsing binaries (EBs) have always been an important tool for testing and
determining the physical properties of stars
\citep{A2008021,A2008029,Torres09}. They are composed of two stars that, when
orbiting each other, produce periodic eclipses. The great potential of EBs is
that their orbital motion, inferred from the radial velocity curves, and the
shape of eclipses, obtained from the light curves, can be entirely explained by
the gravitation laws and the geometry of the system \citep[see][for
details]{Hilditch01}.

The direct determination of the radii of the components of EB systems made that
several authors \citep[e.g.,][]{A2008024,A2008020} suggested the possibility of
using EBs for deriving distances. The only additional requirement to determine
the absolute luminosity of an EB system and, hence the distance, is the surface
brightness or, equivalently, the effective temperature of the components.

The potential of EBs to derive distances encouraged several projects to obtain
direct distance determinations, either within the Milky Way
\citep[e.g.,][]{A2009043}, the Magellanic Clouds \citep[e.g.,][and references
therein]{A2003004}, or M\,31/M\,33 \citep[DIRECT project,][]{A2006018}. In
1999, a new project was started to obtain a direct distance determination to
the Andromeda Galaxy (M\,31) from EBs \citep{A2003018,P2004001}, providing the
first direct distance determination in \citet[][hereafter
\citetalias{P2005002}]{P2005002}. 

The main interest for an accurate distance determination to M\,31 lies on the
potential of this galaxy to be a first-class distance calibrator
\citep[e.g.,][]{A2006008}. The reasons for this are: {\em (1)} Contrary to the
Magellanic Clouds, the distance to M\,31 is large enough so that its geometry
does not introduce any systematics in the final distance determination; {\em
(2)} typically with a moderate reddening value
\citep[$E(B-V)=0.16\pm0.01$,][]{A2009002}, it is close enough to enable the
individual identification of stars suitable for distance determination (such as
EBs or Cepheids); {\em (3)} an Sb I--II giant spiral galaxy (like M\,31)
provides an appropriate local counterpart for the galaxies commonly used for
distance determination \citep[e.g.,][]{A2005021}; and {\em (4)} M\,31 can also
provide an absolute calibration of the Tully-Fisher relationship, enabling the
calibration of the furthest distance determination methods. Therefore, the
characteristics of this spiral galaxy make it an important step of the cosmic
distance scale.

As mentioned above, our project already provided the first direct distance
determination to M\,31 \citepalias{P2005002}. However, the excellent dataset
obtained (Sect.~\ref{obsdistm31}) allowed the determination of additional
distances in order to further constrain the distance to M\,31. Therefore, in
the present work, we are presenting the second direct distance determination to
M\,31 from an EB system (Sects.~\ref{ssrvsb2b}--\ref{ssevosb2b}). This result
(the last one until further spectroscopic observations can be secured) enables
a critical discussion on the distance to M\,31 (Sect.~\ref{sdistm31}). 

\section{Observations}
\label{obsdistm31}

Photometric time series (in $B$ and $V$ passbands) were acquired with the Wide
Field Camera at the 2.5 m Isaac Newton Telescope at La Palma (Spain). The field
of observation covers $34'\times34'$ at the North-Eastern part of M\,31. Around
260 images were obtained in each passband during five campaigns between 1999
and 2003. The Difference Image Analysis technique \citep[DIA,][]{A2002001} was
used to perform the photometric data reduction \citep[thoroughly explained
in][hereafter \citetalias{P2006002}]{P2006002}, providing light curves for more
than 3\,964 variable stars, with 437 being identified as EBs. In addition, the
$V$ photometric observations of the DIRECT project \citep{A2006020} were also
considered. Although the observations are somewhat noisier, they allowed us to
extend the time baseline to 1996 and to ensure the consistency of our DIA
photometry. 

The EB sample was inspected to detect those targets more suitable for distance
determination \citepalias[see][]{P2006002}. A $5'\times5'$ region containing
five of the selected EBs was observed with the multi-object spectrograph (GMOS)
at the Gemini-North telescope (program ID GN-2004B-Q-9). The instrumental setup
was set to provide the highest possible resolution (R=3\,744) using the slit
width of 0\,\farcs5 and the B1200\_G5301 grism. The spectra cover the wavelength
interval between 390 and 530 nm, with two gaps between 436.7--436.9 and
485.8--487.0 nm. A total of eight exposures were taken between September and
November 2004 with an integration time of 4100 seconds each and an additional
one with an exposure time of 3240 seconds in February 2005
(Table~\ref{trvsb2b}). The data reduction was performed with the Gemini IRAF
package version 1.7, providing spectra with a S/N between 6 and 39 for the five
selected EBs.

\begin{table*}[tb]
 \caption{Log of spectroscopic observations and heliocentric radial
velocity determinations (when available).}
 \label{trvsb2b}
 \centering
 \begin{tabular}{lrccccr@{$\pm$}lr@{$\pm$}l}
  \hline
  \hline
  Date & Time & Exp. time & S/N & HJD  & Phase & \multicolumn{2}{c}{Primary} & \multicolumn{2}{c}{Secondary} \\
  & $[$UT$]$& $[$s$]$ & & & & \multicolumn{2}{c}{$[$km~s$^{-1}$$]$} & \multicolumn{2}{c}{$[$km~s$^{-1}$$]$} \\
  \hline
  Nov 7, 2004 & 10:14 & 4100 & 22 & 2\,453\,316.931 & 0.272 & -370.1 & 8.2 & 143.5 & 30.3 \\
  Sep 15, 2004 & 13:29 & 4100 & 11 & 2\,453\,264.066 & 0.467 & \multicolumn{2}{c}{\ldots} & \multicolumn{2}{c}{\ldots}\\
  Feb 12, 2005 & 5:39 & 3240 & 8 & 2\,453\,413.734 & 0.524 & \multicolumn{2}{c}{\ldots} & \multicolumn{2}{c}{\ldots}\\
  Nov 12, 2004 & 7:35 & 4100 & 15 & 2\,453\,321.820 & 0.659 & 14.3 & 13.9 & -414.1 & 25.5 \\
  Nov 12, 2004 & 8:49 & 4100 & 16 & 2\,453\,321.872 & 0.684 & 57.3 & 13.5 & -448.2 & 24.8 \\
  Nov 10, 2004 & 9:00 & 4100 & 24 & 2\,453\,319.879 & 0.711 & 30.6 & 9.9 & -455.5 & 17.4 \\
  Sep 14, 2004 & 8:53 & 4100 & 20 & 2\,453\,262.874 & 0.885 & \multicolumn{2}{c}{\ldots} & \multicolumn{2}{c}{\ldots}\\
  Sep 12, 2004 & 9:45 & 4100 & 13 & 2\,453\,260.910 & 0.927 & -69.7 & 18.1 & -309.2 & 19.4 \\
  Oct 17, 2004 & 6:31 & 4100 & 14 & 2\,453\,295.776 & 0.946 & -58.1 & 24.2 & -271.7 & 22.9 \\
  \hline
 \end{tabular}
\end{table*}

After careful analysis of the five EBs, two of the observed targets resulted to
be optimum for obtaining a precise distance determination to M\,31:
\object{M31V\,J00443799+4129236} and M31V\,J00443610+4129194. The first target
was already analyzed in \citetalias{P2005002} and the analysis of the second
target is presented here. The three remaining EBs are unsuitable for distance
determination. One of them is a single-line EB and, therefore, the absolute
properties of the stars cannot be obtained. Another EB is too faint for an
accurate analysis of its spectra. The third one is a triple-line system, where
the important contribution of the third light prevents a precise distance
determination. 

\section{Radial velocities}
\label{ssrvsb2b}

The determination of radial velocities (RVs) was performed, as in the case of
\citetalias{P2005002}, with TODCOR \citep{A2008019} and the
ATLAS9\footnote{Available at: \texttt{http://kurucz.cfa.harvard.edu/}} and
TLUSTY\footnote{Available at: \texttt{http://nova.astro.umd.edu/}}
\citep{A2009040,Lanz07} synthetic models. In this case, the best pair of
synthetic spectra was determined with the following iterative approach. As a
first step, the preliminary \citet[][hereafter \citetalias{A2005013}]{A2005013}
fit (Sect.~\ref{sswdsb2b}) was used to define an initial list with pairs of
models having a temperature ratio, gravity ratio and rotational velocities
compatible with the \citetalias{A2005013} parameters. All the model pairs in
the initial list were then used to determine RVs and a simple RV curve model
was fitted to the obtained values. The free parameters in the fit were the
semi-major axis ($a$), the systemic velocity ($\gamma$) and the mass ratio
($q$), while the period and reference time was fixed from the
\citetalias{A2005013} solution. From all the derived solutions, the models
having a lower dispersion around the fitted RV curve were selected and all the
neighboring models in the space of parameters were also attempted. The process
was repeated until the local minimum was found, (i.e., none of the neighboring
models has a scatter lower than the selected pair of models). 

Since the resulting RVs depend, to some extent, on the \citetalias{A2005013}
fit, the obtained values were used to find a new solution with the
\citetalias{A2005013} (Sect.~\ref{sswdsb2b}). The resulting
\citetalias{A2005013} solution was then used to determine new RVs. The process
was repeated until the pair of synthetic models providing the best fit was the
same in two iterations. The final solution shown in Table~\ref{trvsb2b}
contains all the RVs (corrected to the heliocentric reference frame) that
remained after the \citetalias{A2005013} fit (with $3\sigma$ clipping).
Rejected observations shown in Fig.~\ref{sb2b} were obtained during eclipses
(one of them corresponds to the spectrum of February 2005 with a shorter
exposure time). 

The rejection of the observation at phase 0.885 cannot be explained neither by
the proximity to the nodes (since other observations are obtained at larger
phases) nor by a lower S/N of the observed spectrum. After some additional
tests, it was observed that the rejected observation could be recovered with
different pairs of templates at the cost of losing other observations close to
the nodes and a larger dispersion of the fit. In all the different solutions
attempted, the systemic velocities and semi-amplitudes were in perfect
agreement with the solution presented here.

\section{Mass and radius determination}
\label{sswdsb2b}

The mass and radius determination was performed following the fitting procedure
described in \citetalias{P2005002}. Both semi-detached configurations (with
either the primary or the secondary filling the Roche lobe) yield fits with
the same residuals. The final configuration was adopted after the temperature
determination analysis (Sect.~\ref{ssteffsb2b}). The spectra resulting from the
disentangling indicate that the secondary is much fainter than the primary,
while the assumption of the secondary filling its Roche lobe yields a flux
ratio of $F_{V,S}/F_{V,P}=0.85$. Therefore, a configuration where the primary
(instead of the secondary) fills the Roche lobe was used in this case. 

In order to ensure the viability of the adopted scenario, and to rule out any
other possible configurations, additional fits adopting a detached
configuration were performed. The resulting parameters revealed that, depending
on the initial value of the surface potential of the secondary component,
either the primary or the secondary tended to fill their respective Roche
lobes. 

\begin{figure}
 \centering
 \resizebox{\hsize}{!}{\includegraphics*{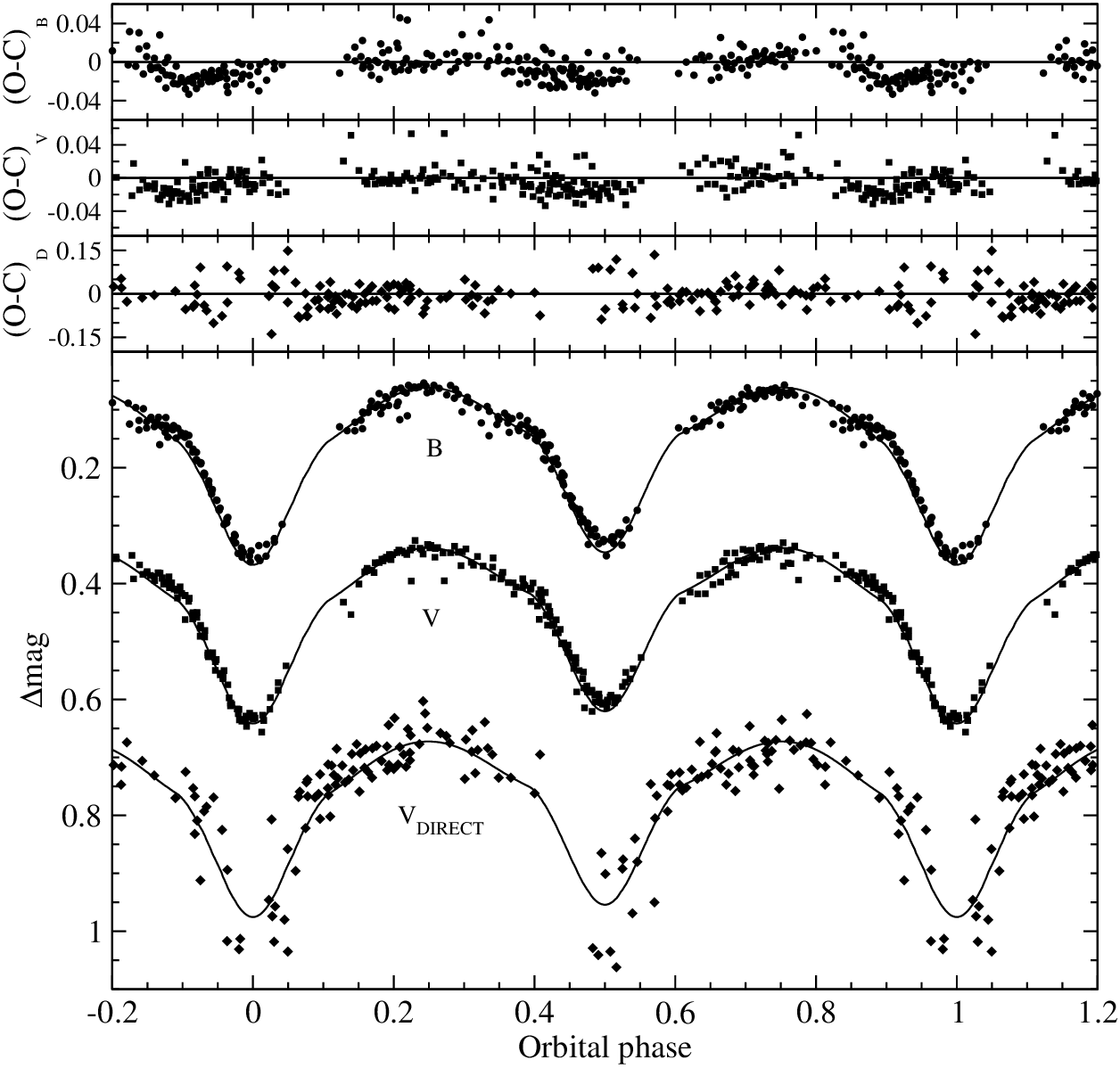}}
 \resizebox{\hsize}{!}{\includegraphics*{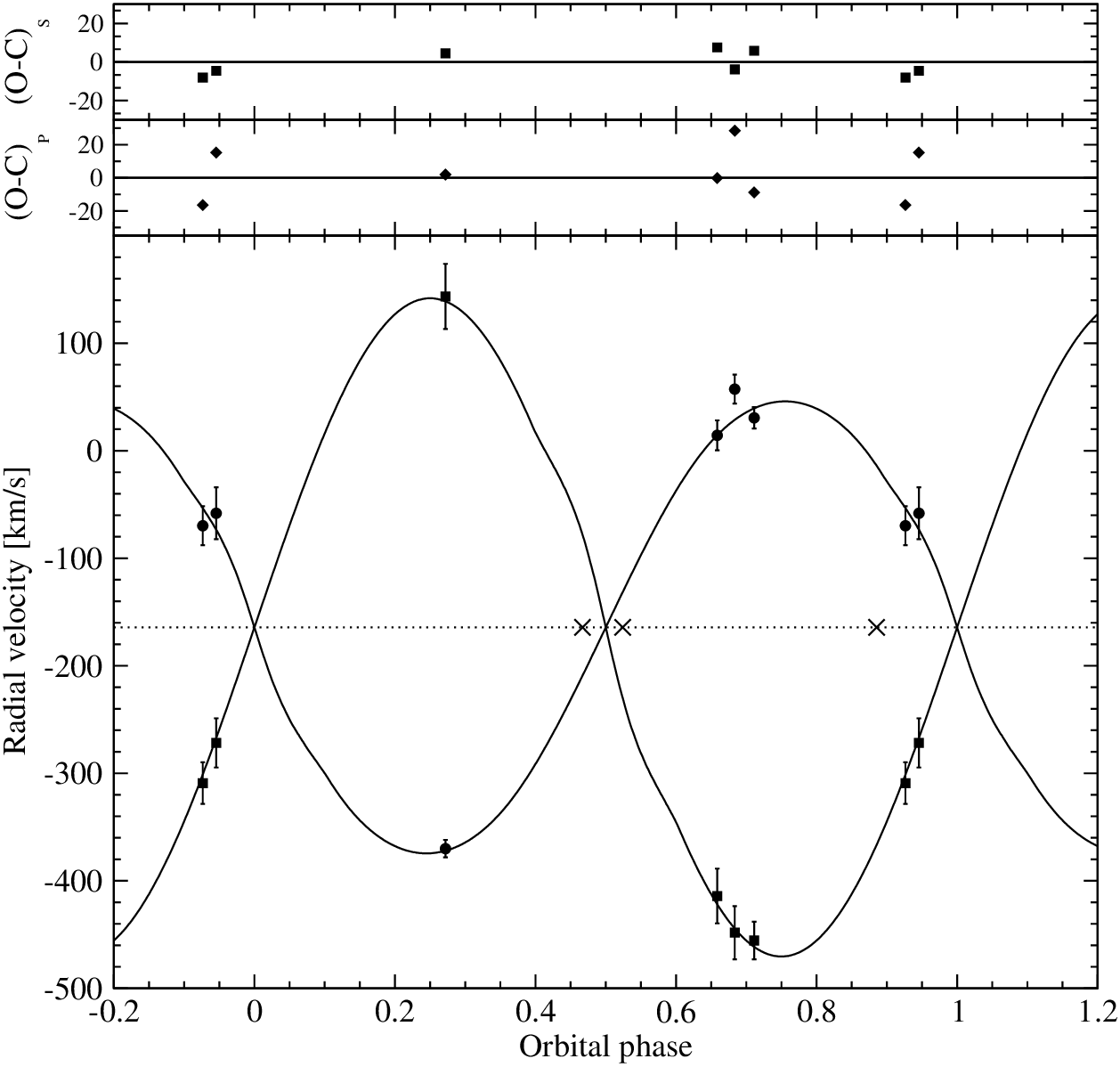}}
 \caption{Observations for M31V\,J00443610+4129194 and corresponding
\protect\citetalias{A2005013} fits. \emph{Top:} Light curve fits and
corresponding residuals. \emph{Bottom:} RV curve fits with RVs for the primary
(circles) and secondary (squares) components with corresponding residuals. The
phases of rejected RV observations are also indicated (crosses).}
 \label{sb2b}
\end{figure}

With the adopted configuration, the final rms residuals are 0.014~mag in $B$,
0.015~mag in $V$, and 0.047~mag in the DIRECT $V$ light curve. The residuals of
the RVs are 13 and 6 km~s$^{-1}$ for the primary and secondary components,
respectively. The light and RV curves, with their respective fits superimposed,
are shown in Fig.~\ref{sb2b}. The resulting best-fitting elements listed in
Table~\ref{twdsb2b} reveal two components with masses and radii of
$\mathcal{M}_P=21.7\pm1.7$ M$_\odot$, $R_P=9.2\pm0.2$ R$_\odot$ and
$\mathcal{M}_S=15.4\pm1.2$ M$_\odot$, $R_S=5.6\pm0.4$ R$_\odot$. 

\begin{table}[tb] 
 \caption{Fundamental properties of M31V\,J00443610+4129194 derived from the
analysis with \protect\citetalias{A2005013}.}
 \label{twdsb2b}
 \centering
 \begin{tabular}{lr@{$\pm$}lr@{$\pm$}l}
  \hline
  \hline
  System properties & \multicolumn{4}{c}{ } \\
  \hline
  $B_{\rm max}^a$ (mag) & \multicolumn{2}{r@{$\pm$}}{19.832} & \multicolumn{2}{@{}l}{0.013} \\
  $V_{\rm max}^a$ (mag) & \multicolumn{2}{r@{$\pm$}}{19.948} & \multicolumn{2}{@{}l}{0.015} \\
  $P$ (days) & \multicolumn{2}{r@{$\pm$}}{2.048644} & \multicolumn{2}{@{}l}{0.000007} \\
  $t_{\rm min}$ (HJD) & \multicolumn{2}{r@{$\pm$}}{2\,452\,908.694} & \multicolumn{2}{@{}l}{0.004} \\
  $i$ (deg) & \multicolumn{2}{r@{$\pm$}}{69.9} & \multicolumn{2}{@{}l}{1.6} \\
  $\gamma$ (km~s$^{-1}$) & \multicolumn{2}{r@{$\pm$}}{-164} & \multicolumn{2}{@{}l}{5} \\
  $a$ (R$_\odot$) & \multicolumn{2}{r@{$\pm$}}{22.6} & \multicolumn{2}{@{}l}{0.5} \\
  $q\equiv\mathcal{M}_S/\mathcal{M}_P$ & \multicolumn{2}{r@{$\pm$}}{0.71} & \multicolumn{2}{@{}l}{0.04} \\
  $T_{\mathrm{eff},S}/T_{\mathrm{eff},P}$ & \multicolumn{2}{r@{$\pm$}}{0.897} & \multicolumn{2}{@{}l}{0.019} \\
  $F_{B,S}/F_{B,P}^a$ & \multicolumn{2}{r@{$\pm$}}{0.33} & \multicolumn{2}{@{}l}{0.03} \\
  $F_{V,S}/F_{V,P}^a$ & \multicolumn{2}{r@{$\pm$}}{0.33} & \multicolumn{2}{@{}l}{0.03} \\
  $F_{D,S}/F_{D,P}^a$ & \multicolumn{2}{r@{$\pm$}}{0.33} & \multicolumn{2}{@{}l}{0.05} \\
  \hline
  Component properties & \multicolumn{2}{c}{Primary} & \multicolumn{2}{c}{Secondary}\\
  \hline
  $R$ (R$_\odot$) & 9.2 & 0.2 & 5.6 & 0.4 \\
  $\mathcal{M}$ (M$_\odot$) & 21.7 & 1.7 & 15.4 & 1.2 \\
 $\log g$ (cgs) & 3.85 & 0.02 & 4.12 & 0.05 \\
  $K^b$ (km~s$^{-1}$) & 210 & 9 & 306 & 11 \\
  $v_{\rm sync}\sin i$ (km~s$^{-1}$) & 213 & 6 & 131 & 8 \\
  \hline
  \multicolumn{5}{l}{$^a$ Out of eclipse average: $\Delta\phi=[0.20-0.30,0.70-0.80]$} \\
  \multicolumn{5}{l}{$^b$ Including non-Keplerian corrections}\\
 \end{tabular}
\end{table}

Light curves measured using DIA photometry require an accurate reference flux
determination to avoid any bias in the scale of the light curves. Following the
same procedure used for the EB in \citetalias{P2005002}, several light-curve
fits were performed with a variable third-light contribution ($l_3$). Again,
the excellent agreement with DIRECT light curve and the convergence of the fits
to $l_3\sim0$, ensures that the flux zero-point is correct and rules out any
possible relevant blends with unresolved companions. 

\section{Temperature and distance determination}
\label{ssteffsb2b}

The last required ingredient to obtain a direct distance determination to M\,31
is the temperature and line-of-sight absorption determination. Again, following
the same procedure as in \citetalias{P2005002}, we determined the temperature
of the components by modeling their disentangled spectra. The spectral
disentangling was performed with the KOREL program \citep{A2008030} with all
the required parameters fixed to the values determined from the
\citetalias{A2005013} analysis (Sect.~\ref{sswdsb2b}).  

\begin{figure*}
 \centering
 \includegraphics[width=180mm]{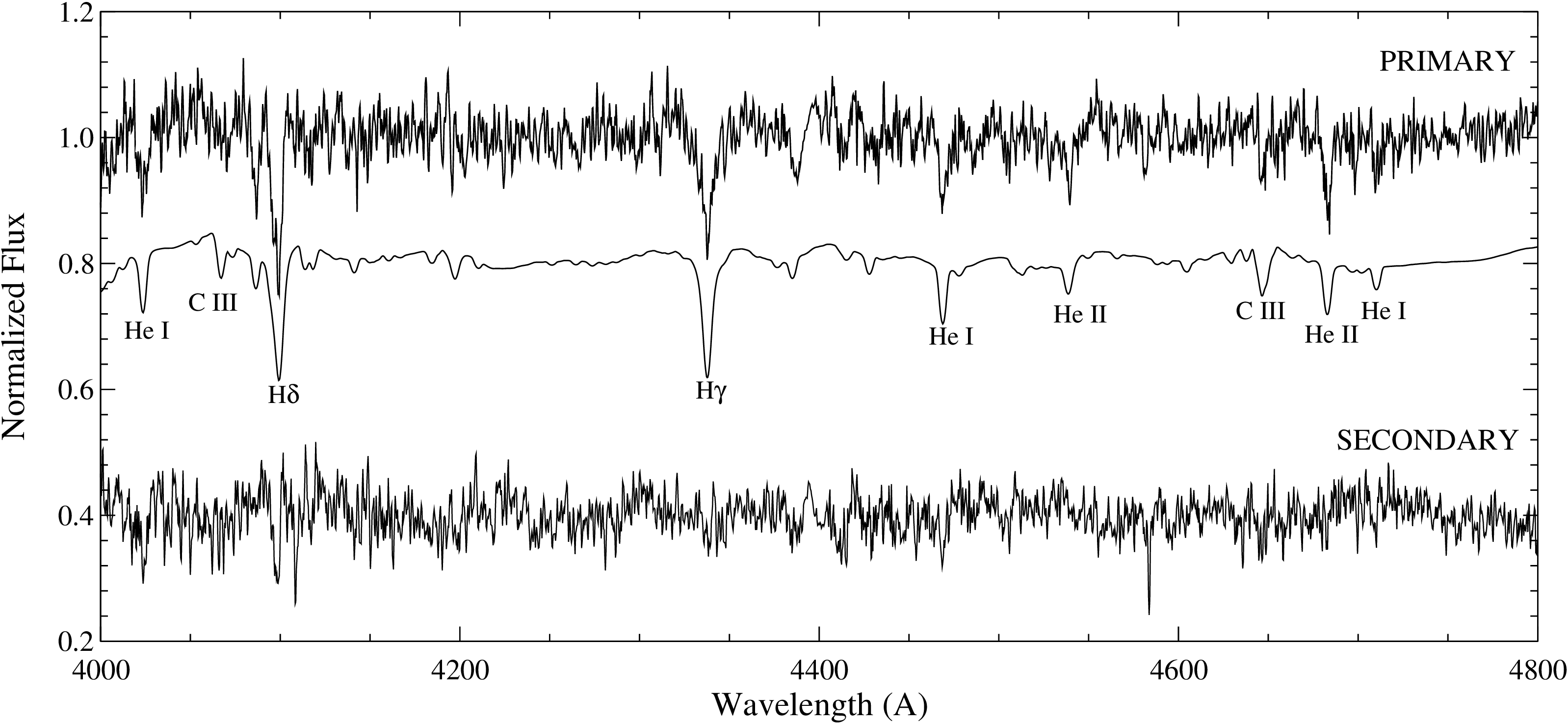}
  \caption{Comparison of the individual disentangled spectrum for the primary
component of M31V\,J00443610+4129194 (above) with the TLUSTY synthetic spectrum
(middle). The bottom spectrum shows the noisy disentangled spectrum of the
secondary component.}
 \label{disspec2b}
\end{figure*}

Considering the low S/N of the spectrum of the secondary, the fitting procedure
was restricted to the primary component. The temperature ($T_\mathrm{eff}$),
surface gravity ($\log g$) and rotational velocity ($v_{\rm rot}\sin i$) were
fitted considering TLUSTY templates. Spectra with solar metallicity were
adopted, since they were found to provide good fits to the observed spectrum.
In addition, the EB analyzed in \citetalias{P2005002}, only 29 arcsec from
M31V\,J00443610+4129194, was found to have solar metallicity. The resulting
values are listed in Table~\ref{tteffsb2b} together with the values of absolute
luminosity ($M_V$) and intrinsic color ($(B-V)_0$). These values were computed
by scaling the absolute magnitude in the models of \citet{A2009028} (for a
given temperature and gravity) with the observed radii of the stars. When
needed, the values resulting from the \citetalias{A2005013} fit were used
(e.g., to determine the temperature of the secondary). This procedure is
equivalent to the one used in \citetalias{P2005002}, where instead of
performing the synthetic photometry to obtain the surface fluxes, they were
obtained from the models. 

Finally, once the absolute magnitudes of the components are determined, the
determination of the distance is straightforward. The resulting distance
modulus to M31V\,J00443610+4129194 is $(m-M)_0=24.30\pm0.11$ mag or,
equivalently, $d=724\pm37$~kpc. 

\begin{table}[tb]
 \caption{Fundamental properties of M31V\,J00443610+4129194 and resulting
distance determination, derived from the modeling of the optical spectrum of
the primary component, using TLUSTY atmosphere models and the values in
Table~\ref{twdsb2b}.}
 \label{tteffsb2b}
 \centering
 \begin{tabular}{lr@{$\pm$}lr@{$\pm$}l}
  \hline
  \hline
  System properties & \multicolumn{4}{c}{ } \\
  \hline
  $M_V$ (mag) & \multicolumn{2}{r@{$\pm$}}{-4.90} & \multicolumn{2}{@{}l}{0.08} \\
  $E(B-V)$ (mag) & \multicolumn{2}{r@{$\pm$}}{0.18} & \multicolumn{2}{@{}l}{0.02} \\
  $A_V$ (mag) & \multicolumn{2}{r@{$\pm$}}{0.55} & \multicolumn{2}{@{}l}{0.08} \\
  $(m-M)_0$ (mag) & \multicolumn{2}{r@{$\pm$}}{24.30} & \multicolumn{2}{@{}l}{0.11} \\
  \hline
  Component properties & \multicolumn{2}{c}{Primary} & \multicolumn{2}{c}{Secondary}\\
  \hline
  $T_\mathrm{eff}$ (K) & 33\,600 & 600 & 30\,100 & 900 \\ 
  $\log g$ (cgs) & 3.86 & 0.12 & \multicolumn{2}{c}{\ldots} \\ 
  $v_{\rm rot}\sin i$ (km~s$^{-1}$) & 189 & 12 & \multicolumn{2}{c}{\ldots} \\ 
  $M_V$ (mag) & -4.59 & 0.07 & -3.38 & 0.12 \\
  $(B-V)_0$ (mag) & -0.295 & 0.002 & -0.286 & 0.004 \\
  \hline
 \end{tabular}
\end{table}

\section{Comparison with stellar evolutionary models}
\label{ssevosb2b}

The final step towards the characterization of M31V\,J00443610+4129194 is the
comparison of the derived physical properties with stellar evolutionary models.
The comparison has been mainly performed with the Geneva models of
\citet{A2009028}, considering solar metallicity. Other models have been
considered \citep{A2008006} without any major variations on the derived
conclusions. It is important to emphasize that the evolutionary tracks are
built for isolated stars. In close binary systems, the evolution of both stars
can be largely modified, with respect to their isolated evolutionary tracks,
when one of the components fills the Roche lobe and mass transfer takes place
\citep[see, e.g.,][]{A2009007}. In addition, due to the effects of mass loss in
massive stars (due to stellar wind), the current derived masses of the
components will be different from the masses at the Zero-Age Main Sequence
(ZAMS). Therefore, we initially fitted different evolutionary tracks in the
mass--radius diagram to derive the mass of each component at the ZAMS
($\mathcal{M}_{P,\rm ZAMS}=22.1$ M$_\odot$, $\mathcal{M}_{S,\rm ZAMS}=15.4$
M$_\odot$).

\begin{figure}
 \centering
 \resizebox{\hsize}{!}{\includegraphics*{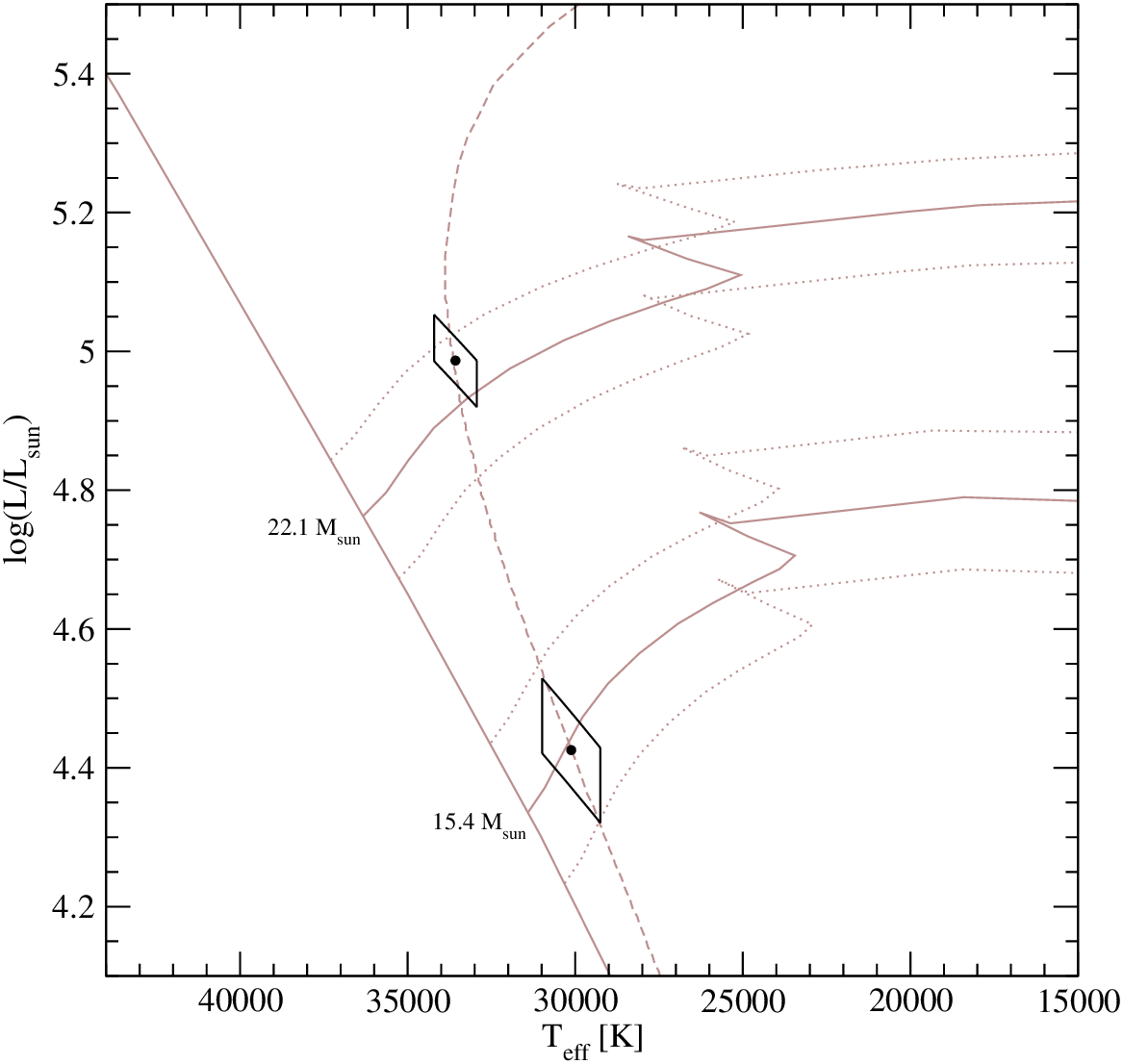}}
 \caption{Comparison of stellar evolutionary models with derived physical
properties of M31V\,J00443610+4129194. Gray solid lines correspond to the ZAMS
and the evolutionary tracks (ZAMS masses are indicated). The best fitting
isochrone of 4.2 Myr is also shown (gray dashed line). Gray dotted lines denote
the uncertainties in the derived masses. The skewed rectangular boxes
correspond to $1\sigma$ error loci.}
 \label{evosb2b}
\end{figure}

Once the evolutionary tracks are adopted, the Hertzsprung-Russell (H-R) diagram
can be studied. The location of the components on the H-R diagram
(Fig.~\ref{evosb2b}) reveals that both agree with their predicted
evolutionary tracks. The derived properties seem characteristic of a detached
system. However, as explained in Sect.~\ref{sswdsb2b}, the
\citetalias{A2005013} solutions supposing a detached configuration converge to
solutions where either the primary or the secondary fill their respective Roche
lobes. In addition, the disentangled spectra seem to favor the case where the
primary is filling the Roche lobe, since it is much brighter than the
secondary. All these observations can be explained supposing that
M31V\,J00443610+4129194 is a pre-mass transfer EB system, where the primary
component is \emph{almost} filling its Roche lobe. 

In a pre-mass transfer EB, both components are detached and basically should
follow the evolution of single stars. This explains why the observed properties
agree with the evolutionary tracks, since each component has evolved without
any major interaction with the other component. This could also explain why the
most massive component can be assumed to be filling the Roche lobe. What is
generally observed in semi-detached EBs is the Algol paradox, where the donor
is the less massive component \citepalias[as in the case of the EB
in][]{P2005002}. The reason of this phenomenon \citep{A2008032} is that, when
the donor is more massive than the companion, the mass transfer is very rapid,
taking place on a thermal time scale ($\sim10^5$~yr). On the contrary, when the
donor is the less massive companion, the mass transfer can last as long as the
main-sequence life-time of the donor. The lack of any signature of intense mass
transfer (such an O'Connell effect in the light curves), combined with the
short timescale of the process, makes that a situation where the most massive
component is filling the Roche lobe is unlikely. However, the time spent in
a situation where the massive component can be almost filling the Roche lobe
has a typical timescale of $\sim0.5$~Myr, making the situation more likely to
be observed. 

The close agreement of the observations with the theoretical models, apart from
clarifying the evolutionary stage of M31V\,J00443610+4129194, also allowed the
fitting of an isochrone, resulting in an estimated coeval age of
$4.2\pm0.4$~Myr. 

\section{The distance to M\,31}
\label{sdistm31}

The two analyzed EBs can be used to provide a distance determination to the
M\,31 galaxy (Table~\ref{tebdist}). As mentioned in \citetalias{P2005002}, the
derived distances to each EB can be associated to the center of M\,31, because
the correction due to the location of the EB within the galaxy is negligible
($\sim0.3$\%). Therefore, the two independently derived distances, in mutual
agreement within their one sigma error bars, show that EBs can be used to
derive precise and accurate distances to M\,31.

\begin{table*}[tb]
 \caption{Distance determinations to M\,31 from EBs.} 
 \label{tebdist}
 \centering
 \begin{tabular}{lr@{$\pm$}lr@{$\pm$}lr@{$\pm$}lr@{$\pm$}lr@{$\pm$}l}
  \hline
  \hline
  EB system & \multicolumn{2}{c}{$V$} & \multicolumn{2}{c}{$M_V$} & \multicolumn{2}{c}{$A_V$} & \multicolumn{2}{c}{$(m-M)_0$} & \multicolumn{2}{c}{Distance} \\
  & \multicolumn{2}{c}{[mag]} & \multicolumn{2}{c}{[mag]} & \multicolumn{2}{c}{[mag]} & \multicolumn{2}{c}{[mag]} & \multicolumn{2}{c}{[kpc]} \\
  \hline
  M31V\,J00443799+4129236 & 19.27 & 0.02 & -5.77 & 0.06 & 0.60 & 0.10 & 24.44 & 0.12 & 772 & 44 \\
  M31V\,J00443610+4129194 & 19.948 & 0.015 & -4.90 & 0.08 & 0.55 & 0.08 & 24.30 & 0.11 & 724 & 37 \\
  \hline
  Weighted mean & \multicolumn{2}{c}{\ldots} & \multicolumn{2}{c}{\ldots} & \multicolumn{2}{c}{\ldots} & 24.36 & 0.08 & 744 & 33 \\
  \hline
 \end{tabular}
\end{table*}

The two double-line EBs have been combined to derive a weighted mean distance
to M\,31 of $744\pm33$~kpc or $(m-M)_0=24.36\pm0.08$ mag. This resulting
distance relies on the modeling of two different EBs and, therefore, can be
considered to be:

\paragraph{Direct.} The distance determination from EBs does not rely on
previous calibrations and, since it is derived using a one-step procedure, it
can be considered direct. Therefore, a recalibration of the distance
to other Local Group galaxies (such as \object{LMC}) or a variation on the
zero-point of any standard candle (such as Cepheids) has no effect on the
derived distance. Furthermore, any standard candle in M\,31 can be calibrated
using our derived distance.

\paragraph{Precise.} The uncertainty on the derived distance modulus of 0.08
mag represents a distance determination with an error of only 4\%, which is
remarkable, given the faintness of the studied targets. Although other distance
determinations to M\,31 with smaller errors have been reported \citepalias[see
a list in][]{P2006002}, most of these results do not consider the effect of
systematics. In fact, the derived uncertainty is equal to the standard
deviation resulting from the combination of the non-direct distance
determinations in \citetalias{P2006002} ($(m-M)_0=24.39\pm0.08$ mag) and,
therefore, our derived distance is equally precise. The uncertainty in the
derived distance could be improved with the analysis of additional EBs
\citepalias[already identified in the list of 24 EBs provided in ][]{P2006002},
resulting in a distance determination to M31 with a relative uncertainty of
2--3\% and free of most systematic errors. This result would represent the most
accurate and reliable distance determination to this important Local Group
galaxy. 

\paragraph{Accurate.} One of the most important points when determining
distances is the effect of possible systematics in the derived value. Contrary
to other distance determinations, the uncertainty in our distance modulus
includes most, if not all, the possible systematics. In particular, the
possible sources of systematic errors, and the corresponding considerations,
can be summarized in the following points:

\begin{itemize} 
\item Photometry. The photometry has been compared with other
catalogs \citepalias{P2006002} and checked to be well below 0.03 mag for the
magnitude of the selected EBs. 

\item Assumed configuration in the modeling of the EBs. The
configuration assumed is completely independent for each one of the two EBs
used for distance determination and results in distances that agree within the
uncertainties. In addition, M31V\,J00443799+4129236 has clear evidences of
being a semi-detached EB (O'Connell effect, no eccentricity, etc.) and
M31V\,J00443610+4129194 has been thoroughly tested for any other possible
configuration, with none of them being capable to reproduce the observations. 

It has traditionally been argued that detached EBs are the only systems capable
to provide accurate distances. The most common reason is that non-detached EBs
are affected by the proximity of the components, introducing distortions and
reflection effects. However, the proximity of the components can properly be
taken into account by current modeling algorithms (such as
\citetalias{A2005013}). In addition, the fact that one of the components fills
the Roche lobe decreases the number of free parameters and greatly improves the
stability of the solution \citep{A2009035}. On the other hand, there are some
other effects, missing in detached EB, that could introduce some systematics in
the solutions (e.g., hot spots, circumstellar disks, etc.). In any case, these
effects can be observed from the acquired data and properly modeled when they
have a mild effect on photometry. A clear example of these effects is the
presence of the O'Connell effect in M31V\,J00443799+4129236. The modeling
performed, with the introduction of a hot spot, enabled the use of
M31V\,J00443799+4129236 for distance determination. Therefore, using
semi-detached EBs reduces the number of free parameters, at the cost of having
to account for some other effects that, when properly considered, should
introduce no systematic error in the derived distances. 

\item Radial velocities. Several aspects have been considered with respect to
the radial velocity determinations. TODCOR is a well tested program that has
been checked to introduce some systematics only for spectra with a narrow
wavelength coverage \citep{A2009034}, which is not our case. Moreover, the
possible systematics introduced by the use of Balmer lines (that can be
affected by stellar winds) is compensated with the incorporation of He lines.
In addition, the stellar wind would introduce a bias mainly in the systemic
velocity, which has no impact in the final distance determination. In any case,
the derived systemic velocity ($\gamma=164\pm5$ km~s$^{-1}$) is in excellent
agreement with the galactic rotation curve of M\,31 \citep[see, e.g.,][where
the studied EB is at $(X,Y)\simeq(23\farcm3 N,8\farcm4 E)$]{A2009047} and very
similar to the systemic velocity of the neighboring EB studied in
\citetalias{P2005002} ($\gamma=173\pm4$ km~s$^{-1}$). Therefore, the effect of
any systematics on the derived radial velocities can be considered negligible. 

\item Stellar atmosphere models used to determine the surface flux. This is
probably the major source of systematic errors. In order to transform the
derived temperatures and gravities to surface fluxes, the synthetic spectra are
convolved with filter responses and converted to absolute magnitudes and colors
using a series of calibrations \citep[see][for a description on the
transformations applied]{A2009028}. The accuracy of these transformations is
usually defined within a few hundredths of magnitude (see \citeauthor{A2008005}
\citeyear{A2008005}, for various determinations of the bolometric corrections
for the Sun and the colors of Vega, and \citeauthor{A2009046}
\citeyear{A2009046}, for the dispersion on the bolometric correction for hot
stars). Therefore, we expect such systematic error to be no larger than a few
per cent in flux, having an effect below 0.06 mag in the distance modulus.

\item Line-of-sight absorption. The determination of the line-of-sight
absorption benefits from the weak temperature dependence of $(B-V)_0$ above
$\sim30000$ K to obtain an $E(B-V)$ value. Therefore, the main source of
uncertainty at optical wavelengths is the determination of the
total-to-selective extinction ratio ($\mathcal{R}_V$). The value of
$\mathcal{R}_V$ has already been considered to have an uncertainty of 10\% and
larger variations are unlikely, as seen from previous statistical analysis
\citep{A2008009}. The large uncertainty in the value of $\mathcal{R}_V$
increases the line-of-sight absorption error.  In fact, uncertainty in the
extinction represents $\sim$50\% of the total error in the distance
determination. Therefore, any further improvements that reduce the uncertainty
in the line-of-sight absorption could potentially improve the distance
determinations presented here.
\end{itemize}

The precision and accuracy on the derived distance to M\,31 enables a detailed
comparison with previous results. The derived distance modulus is in complete
agreement with the mean value obtained from the combination of non-direct
distance determinations in \citetalias{P2006002}. In fact, \emph{all} the 18
distances reported agree with the EB value within $2\sigma$ and prove that the
different distance indicators are all rather well calibrated within their error
bars. 

During the course of our project, Cepheids have also been used to determine a
distance to M\,31 \citep{P2007002}. The derived distance modulus of
$(m-M)_0=24.32\pm0.12$ mag represents an additional distance determination to
M\,31 that is fully compatible with the EB value. The consistency of this
result is specially valuable, because an EB distance to LMC has been used as
reference. Therefore, the Cepheid distance modulus proves that the EB distances
to LMC and to M\,31 are fully compatible, tightening the extragalactic distance
scale. In fact, Cepheids have proved to be extremely useful and robust to
determine distances to LMC using, as reference, direct distance determinations
to other galaxies. \citet{A2006023} used the Cepheids in \object{NGC\,4258},
which has a direct maser-based distance \citep{A2007030}, to determine a
distance modulus to LMC that is almost identical to the EB value. Therefore,
the derived distances using EBs are not only fully compatible among themselves,
but also to other direct distance determination methods.

An exception to the general agreement is the distance determination to
\object{M\,33} with one EB system \citep{A2006011}. The derived distance
modulus is 0.3 mag larger that the Cepheid distance determination of
\citet{A2005021}, implying a distance to LMC that is incompatible with the
remaining distance determinations. Even a maser-based distance to M\,33 was
obtained favoring the short distance to M\,33 \citep{A2009048}, the
associated uncertainty is too large for a firm establishment of the distance.
Undoubtedly, additional distance determinations to M\,33 with EBs will be
crucial to solve the discrepancy. 

EBs have proved to be excellent distance markers and, arguably, the best method
to provide direct distances for all the galaxies in the Local Group. In
addition, the large number of distance indicators observed in M\,31 strengthens
the importance of this galaxy as the main calibrator for extragalactic distance
determinations. In fact, all the distance indicators used to determine a
distance to LMC \citep{A2009024}, can be used, in principle, in M\,31. In
addition, at least two other distance determination methods can be used in
M\,31 (planetary nebulae luminosity function and Tully-Fisher relationship).
Therefore, after the direct distance obtained with EBs, M\,31 may be the best
extragalactic benchmark for distance determinations. 

\begin{acknowledgements}
This program was supported by grant AYA2006-15623-C02-01/02 of the Spanish
Ministerio de Ciencia e Innovaci\'on (MICINN). F.V. acknowledges support from
MICINN through a Consolider-GTC (CSD-2006-00070) fellowship.
\end{acknowledgements}

\bibliographystyle{aa} 
\bibliography{bibliography} 
\end{document}